# Resource Estimation of CGGI and CKKS scheme workloads on FracTLcore® Computing Fabric

Denis Ovchinnikov, Hemant Kavadia, Satya Keerti Chand Kudupudi, Ilya Rempel, Vineet Chadha, Marty Franz, Paul Master, Craig Gentry, Darlene Kindler, Alberto Reyes and Muthu Annamalai[+]

Corresponding author[+]: mannamalai@cornami.com

## Abstract

Cornami $Mx^2$ accelerates of Fully Homomorphic Encryption (FHE) applications, enabled by breakthrough work [1], which are otherwise compute limited. Our processor architecture is based on the systolic array of cores with in-memory compute capability and a network on chip (NoC) processor architecture called the "FracTLcore® compute fabric processor" ($Mx^2$). Here, we describe the work to estimate processor resources to compute workload in CGGI (TFHE-rs) or CKKS scheme during construction of our compiler backend for this architecture [2]. These processors are available for running applications in both the TFHE-rs Boolean scheme and CKKS scheme FHE applications.

## 1. Introduction

Fully Homomorphic Encryption (FHE) is a technique to run computations on encrypted data without decrypting it. There are many applications of FHE, such as Privacy-preserving Machine Learning (PPML), secure voting, etc. However, FHE is very compute intensive requiring hardware acceleration to be commercially viable. The Cornami $Mx^2$ is a FracTLcore compute fabric processor that supports reconfigurable systolic arrays, immediate memory access, pipeline parallelism and quick reprogramming of function – all features not traditionally available in GPUs or CPUs. An introduction to FHE is skipped in this article whereas it is widely covered in earlier literature including [6].

The SCIFR (Secure Computing Interface FRamework) [3] API delivers a standardized, hardware-agnostic interface for deploying industry-standard Fully Homomorphic Encryption (FHE) frameworks on Cornami's scalable FracTLcore computing fabric ($Mx^2$ systolic array processor). By abstracting hardware orchestration, SCIFR enables dynamic tuning of fabric resources to optimize performance versus cost for encrypted data workloads. Cornami hardware enables the creation of topologies to support FHE algorithms, mixed-schema support for hybrid encryption flows, and full-circuit execution directly in hardware—eliminating co-processor overhead for maximum silicon efficiency (pipeline parallelism, SRAM and HBM access at low latency).

Complemented by the TruStream® SDK [4], developers can extend and customize FHE applications to push cryptographic boundaries and integrate novel algorithmic innovations seamlessly. SCIFR's modular design accelerates development cycles, simplifies integration across market segments such as smart contracts, healthcare, manufacturing, and FinTech, and scales throughput to meet stringent application requirements. Cornami's SCIFR API empowers technical teams to harness high-performance homomorphic encryption at cloud scale securely.

In this article we propose methods to estimate the Cornami's $Mx^2$ resource consumption for FHE applications in both CGGI and CKKS schemes; we need to solve this problem enroute to building our general-purpose compiler for $Mx^2$ architecture. Our work is differentiated from FPGA resource estimation [8] by specifically being tuned to the $Mx^2$ processor, as well as the integration into a compiler pipeline.

## 2. Application Resource Usage on $Mx^2$ for FHE Acceleration

Our software stack, in Fig. 1, provides APIs for both FHE schemes (CGGI, CKKS) to be efficiently executed on the $Mx^2$ chip.

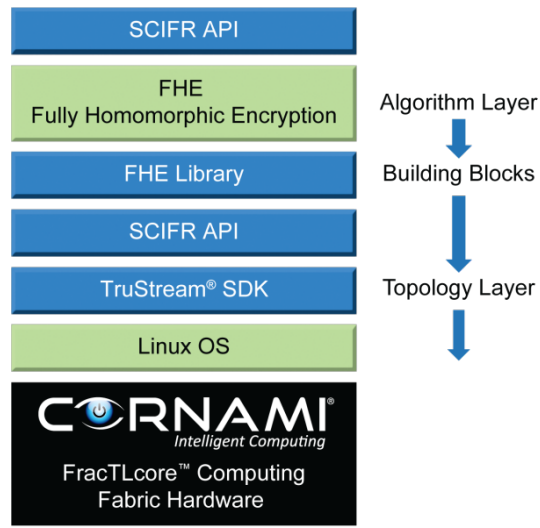

Fig. 1: Cornami SCIFR API stack for building FHE applications on Cornami MX2 processor.

However, in the interest of generality we would like a compiler solution to run FHE workloads on the $Mx^2$ without manual coding of the application in SCIFR or TruStream APIs. For this reason, we pursue a compiler, Fig. 2, for both schemes.

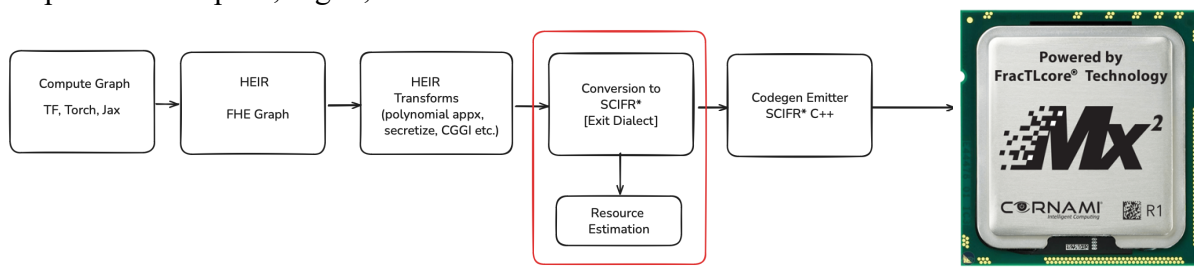

Fig. 2: Cornami compiler backend using the HEIR MLIR pipeline for TFHE-rs/CGGI scheme using the Cornami SCIFR Bool dialect; the compiler backend for CKKS scheme uses a corresponding SCIFRCkks dialect.

HEIR FHE Compiler [6-7] uses the MLIR framework, as a flexible and extensible compiler infrastructure for various modern hardware architectures. HEIR aims to standardize IRs for FHE by supporting interoperability across different FHE schemes such as BGV, CKKS, TFHE-rs, etc.

We build our compiler as a backend on the HEIR compiler targeted toward the $Mx^2$ Cornami chip where CGGI exits to SCIFRBool and CKKS exist to the SCIFRCkks dialect respectively.

Resource estimate gives use (Mx$^2$) *Boards*, *Chips*, *Tiles*, *FCs* (FractalCores the PE), *HBM* and *DDR* memory units. Resource Estimation (FCs) for the compute graphs provides the cumulative count of the operators in FHE scheme as lowered by the compiler into the Cornami compute topologies. For example, the CKKS scheme operators Key Switch (KS), Bootstrap (PBS), Ciphertext Mul, Plaintext Mul, Square, and Rotate are lowered into respective topologies from SCIFRCkks library; to compute the resource required for computing the operator graph on Mx$^2$ chip we use a cumulative sum approach during a graph traversal of the compute graph.

Additionally, we have provided a critical-path (CP) estimation algorithm which gives the estimate of the longest time for input to propagate to the output (time-to-first output ciphertext) for the given circuit. This is relevant since the Cornami architecture and other dataflow machines that are inherently pipelined and smaller critical paths are Pareto optimal in such cases.

**Critical Path Estimation Algorithm**

**Algorithm: ApproximateCriticalPathEstimation**[ $G_{in}$ : Input MLIR Operator Graph]
```
    result = topologicalSort( Gin)
    sinks = getSinks(Gin)
    sources = getSources(Gin)
    cp_result = []
    for node in result[:-1]:
          if node in sources or node in sinks: continue
          cp_result.append( node )
    return cp_result
```

**Algorithm: ExactCriticalPathEstimation**[ $G_{in}$ : Input MLIR Operator Graph]
```
    result = topologicalSort( Gin)
    sinks = getSinks(Gin)
    sources = getSources(Gin)
    cp_result = []
    plen = 0
    for src in sources:
        for sink in sinks:
            if ( src == sink ) continue
            path = getShortestPath(src,sink)
            currlen = len(path)
            if currlen > plen:
                plen  = currlen
                cp_result = path
    return cp_result
```

Code. 1: Algorithm for approximate and exact CP estimation

## 3. Estimation for CGGI Applications

Our compiler Backend for CGGI/TFHE consists of SCIFR Bool MLIR Dialect and Emitter on HEIR framework. We lower CGGI/TFHE Dialects to SCIFR Bool Dialect and generate C++ code on Cornami Concrete/AppStream Framework and run the Generated Code on Emulator and Hardware. We have designed SCIFR Bool as an Exit Dialect for Boolean FHE schemes to model the Boolean gates on Cornami's FracTLCores® computing fabric processor. The currently supported operations and types of SCIFR Bool dialect are shown below:

| Operations | And, Nand, Nor, Or, Xor, XNor, Not, Packed, Lut2, Lut3, LutLinComb, MultiLutLinComb |
|---|---|
| Types | LWE Ciphertext, BootstrapKeyStandard, KeySwitchKey, ServerParameters |

Table. 1: CGGI Dialect Operators and Types (2-bit or 1-bit encrypted inputs)

To develop the estimates of the CGGI elemental operators to run on $Mx^2$ we developed the follow atomic estimates and aggregate these over the compute graph using the MLIR analysis pass pipeline.

| CGGI Op | BS | KS | FC Cores |
|---|---|---|---|
| AndOp | Y | Y | PBS + KS + const |
| NandOp | Y | Y | PBS + KS + const |
| NorOp | Y | Y | *-ibid-* |
| OrOp | Y | Y | *-ibid-* |
| XorOp | Y | Y | *-ibid-* |
| XNorOp | Y | Y | *-ibid-* |
| NotOp. | N | N | 4 to 16 FCs |
| PackedOp | Y | Y | PBS + KS + const |
| Lut2Op | Y | Y | *-ibid-* |
| Lut3Op | Y | Y | *-ibid-* |
| LutLinCombOp | Y | Y | *-ibid-* |
| MultiLutLinCombOp | Y | Y | *-ibid-* |

Table. 2: Estimates of Boolean Operators

We have added MLIR transformations such as a transformation from CGGI to SCIFR Bool and a transformation to add Sections in SCIFR Bool to divide large Boolean circuits into sections that can be mapped to $Mx^2$ processor in a time-division multiplexing fashion. Any other operations, types or transformations can be easily added to our framework as it is extensible and customizable. We have also built an Emitter for C++ code Generation using Concrete/AppStream for the $Mx^2$ processor. This emitter is also extensible, and new emitters can be added in our framework for code generation for other hardware or APIs. For example, the following command on our MLIR tool allows CGGI resource estimation.

```
$ scifr-opt --canonicalize --cggi-tigris-estimator lut_canonicalize.mlir
```

| Operator | LUT Canonicalize | AND Gate | Half Adder | 8-bit Multiply |
|---|---|---|---|---|
| AndOp (FCs) | 0 | 1024 | 256 | 11264 |
| Lut2Op (FCs) | 256 | 0 | 0 | 0 |
| Lut3Op (FCs) | 256 | 0 | 0 | 0 |
| LutLinCombOp (FCs) | 0 | 0 | 0 | 0 |
| MultiLutLinCombOp (FCs) | 0 | 0 | 0 | 0 |
| NandOp (FCs) | 0 | 0 | 0 | 11264 |
| NorOp (FCs) | 0 | 0 | 0 | 0 |
| NotOp (FCs) | 0 | 0 | 0 | 0 |
| OrOp (FCs) | 0 | 0 | 0 | 0 |
| PackedOp (FCs) | 0 | 0 | 0 | 0 |
| XNorOp (FCs) | 0 | 0 | 0 | 4608 |
| XorOp (FCs) | 0 | 0 | 256 | 8960 |
| Total $Mx^2$ Chips | 1 | 1 | 1 | 18 |
| Boards $Mx^8$ | 1 | 1 | 1 | 5 |
| Normalized Critical Path Latency (Cycles) | 1 | 1 | 0.5 | 35 |
| Normalized Time Estimate (unit-time) for 1000-batch size | 1 | 2 | 0.02 | 70.5 |

Table. 3: Sample FHE applications in CGGI mapped to Cornami hardware; we assume time to first output generation as critical path as outlined earlier in the article as well.

## 4. Estimation for CKKS Applications

| **Operations** | AddOp, AddPlainOp, SubOp, SubPlainOp, MulOp, MulPlainOp, RotateOp, ExtractOp, NegateOp, RelinearizeOp, RescaleOp |
|---|---|

Table. 4: Cornami CKKS Dialect Operators

| Applications | CKKS Ops | Critical Path CKKS Ops |
|---|---|---|
| box_blur_4x4 | 40 | 28 |
| box_blur_64x64 | 40 | 28 |
| dot_product_8 | 34 | 24 |
| `gx_kernel_64x64` | 38 | 22 |
| `gx_kernel_8x8` | 38 | 22 |
| `hamming_distance` | 32 | 22 |
| `linear_polynomial_64` | 24 | 16 |
| `quadratic_polynomial` | 32 | 30 |
| `roberts_cross_4x4` | 30 | 18 |
| `roberts_cross_64x64` | 32 | 18 |
| `simple_sum` | 36 | 26 |

Table. 5: CKKS applications lowered into Cornami CKKS Operators; each operator consists of 500+ FCs and the full topology of implementation may consume several Mx$^2$ chips or Mx$^8$ boards (4 Mx$^2$).

To obtain resource estimation of the compute graph, in FHE CGGI or CKKS scheme, we write an Analysis Pass in HEIR just in the same way as for CGGI estimation in section 3. We do resource Estimation (FCs) by Function as cumulative sum of elementary resource consumption of each elemental ops like *Key Switch, Bootstrap, Ciphertext Mul, Plaintext Mul, Square, Rotate* for the CKKS scheme with few additional operators to be completed. Estimates are based on topology created for these operators as in the SCIFRCkks operators. For example, the following command on our MLIR tool allows CKKS resource estimation.

```
$ scifr-opt --ckks-tigris-estimate ckks_example.mlir
```

The performance can be improved by implementing denser PBS/KS operations, Fast-bootstrapping and SIMD. We are working to integrate our block-placer solution for maximizing the available compute cores in one Mx$^2$ chip.

## 5. Results

We have also run experiments to estimate performance of Integer multiplications using our framework. The steps to generate code for integer multiplication are as follows:

1. Write source program in Arith dialect with Plaintext inputs
2. Use `heir-opt` to generate encrypted version (CGGI) with LWE Ciphertext inputs
3. Use `scifr-opt` to convert the CGGI to SCIFR Bool
4. Use `scifr-translate` to generate C++ code with Cornami TFHE-rs implementation

Our current implementation uses a linear combination implementation of various CGGI gate operators using fundamental AND gate LUT. A unique LUT for all 2 encrypted bit gates will reduce the area footprint of the fundamental blocks.

The generated code for `int8` multiplication generates a Boolean circuit with a critical path of 14 Boolean gates deep. According to our estimation a 1 unit time/gate operation gives about 14 unit time per `int8` product on $Mx^2$ with one $Mx^8$ board at occupancy of 50% per chip due to current manual organization of PBS+KS; an automatic place and route solution, (in the works), will improve our utilization closer to 75-90%. The pipeline parallelism of our architecture gains throughput. Batch-size divided by latency 14-unit time/output, gives a throughput of 71 encrypted values generated at a batch size of 1000.

Our current implementation uses a linear combination implementation of various CGGI gate operators using fundamental AND gate LUT. A unique LUT for all 2 encrypted bit gates will reduce the area footprint of the fundamental blocks.

## 6. Conclusion

In this work we estimated computational resources of FHE applications, in both CGGI and CKKS schemes, in preparation to run on Cornami $Mx^2$ architecture. This effort supports our compiler [2] for Boolean and CKKS schemes to provide software toolchain for $Mx^2$ architecture.